\documentclass{mem}
\usepackage{natbib}\usepackage{txfonts}\usepackage{balance}
\usepackage{graphicx}
\usepackage[a4paper,breaklinks,dvipdfm]{hyperref}
\idline{75}{282}
\begin{document}
\def\teff{$T\rm_{eff }$}
\def\kms{$\mathrm {km s}^{-1}$}

\title{
Formation of galactic disks through gas-rich mergers
}

   \subtitle{}

\author{
F. Hammer\inst{1} on behalf of the IMAGES team 
         }
 
  \offprints{F. Hammer}
 
\institute{
GEPI, Observatoire de Paris, CNRS, 92190 Meudon, France
\email{francois.hammer@obspm.fr}
}

\authorrunning{Hammer}

\titlerunning{rebuilding disks after mergers}

\abstract{To probe the progenitors of the numerous massive spirals requires to dissect distant galaxy properties through spatially-resolved kinematics, detailed morphologies and photometry from UV to mid-IR. So far IMAGES is the only representative sample studied that way. Six billion years ago, 50\% of spiral progenitors were experiencing major mergers, evidenced by the combination of their peculiar kinematics and morphology. IMAGES provides the observational point of the spiral rebuilding scenario, which agrees with predictions from the $\Lambda$CDM. It reconciles the disk formation or survival with the observed merger rate and allows to reproduce realistic galactic disks with sufficient angular momentum. Several consequences are expected in the Local Universe, because ancient major mergers had let imprints in galaxy haloes, including the most spectacular cases of NGC5907 and M31. An ancient merger in the later galaxy may have left many debris within the Local Group.
}
\maketitle{}
\begin{quotation}
{\it To Tom Dunnill, his life \& his ideas}
\end{quotation}

\section{Introduction}
Seventy two percent of galaxies with stellar mass larger than 2 $10^{10} M_{\odot}$ are dominated by their disk and spiral structures. The formation of giant galactic disks is directly linked to the origin of their supporting angular momentum, and the orbital angular momentum from major mergers may solve the spin ÒcatastropheÓ \citep{Maller02}. 

This may lead to a novel channel to form large disks, because gas-rich mergers are rebuilding spiral galaxies with 0.06$<$B/T$<$0.5 for 5:1 to 1:1 merger with at least 40\% of gas richness \citep[see also Figure 1]{Hopkins10}. Such gas richness were quite common in the distant Universe \citep{Rodrigues12}. For example, at z$\sim$ 2, most galaxies within a mass range consistent to be progenitors of present-day spirals have $f_{gas}>$ 40\% \citep{Hammer09}. 

Expectations that merger of disks produce ellipticals has been increasingly acknowledged since \cite{Toomre72}. Spiral disk formation from an early-on tidal torque process has shown many weaknesses, from the spin catastrophe \citep{Steinmetz99} to the inability to recover the observed fractions of compact and strong star forming distant galaxies \citep{Hammer05}. It is now challenged because many disks can be rebuilt after gas-rich mergers. Nowadays, the main question is how many and which giant spirals have been formed this way during the last  ten billion years. Other mechanisms have to be analyzed such as cold gas streams or minor mergers. They have to explain the large angular momentum stabilizing giant spiral galaxies, which would exclude numerous minor mergers during a galaxy lifetime, if those had not aligned orbital motions. Simulations show cold-gas streams containing clumps, which are merging galaxies according to \cite{Danovich12}.

This short paper reviews a methodological approach to investigate the emergence of the giant spiral galaxies by studying their progenitors, i.e., galaxies having emitted their light 5 to 8 billion years ago. It is illustrated by the Intermediate MAss Galaxies Evolutionary Sequence (IMAGES) that is the most accomplished survey of distant galaxy physical properties, including spatially resolved kinematics. 

\begin{figure*}[t!]
\resizebox{\hsize}{!}{\includegraphics[clip=true]{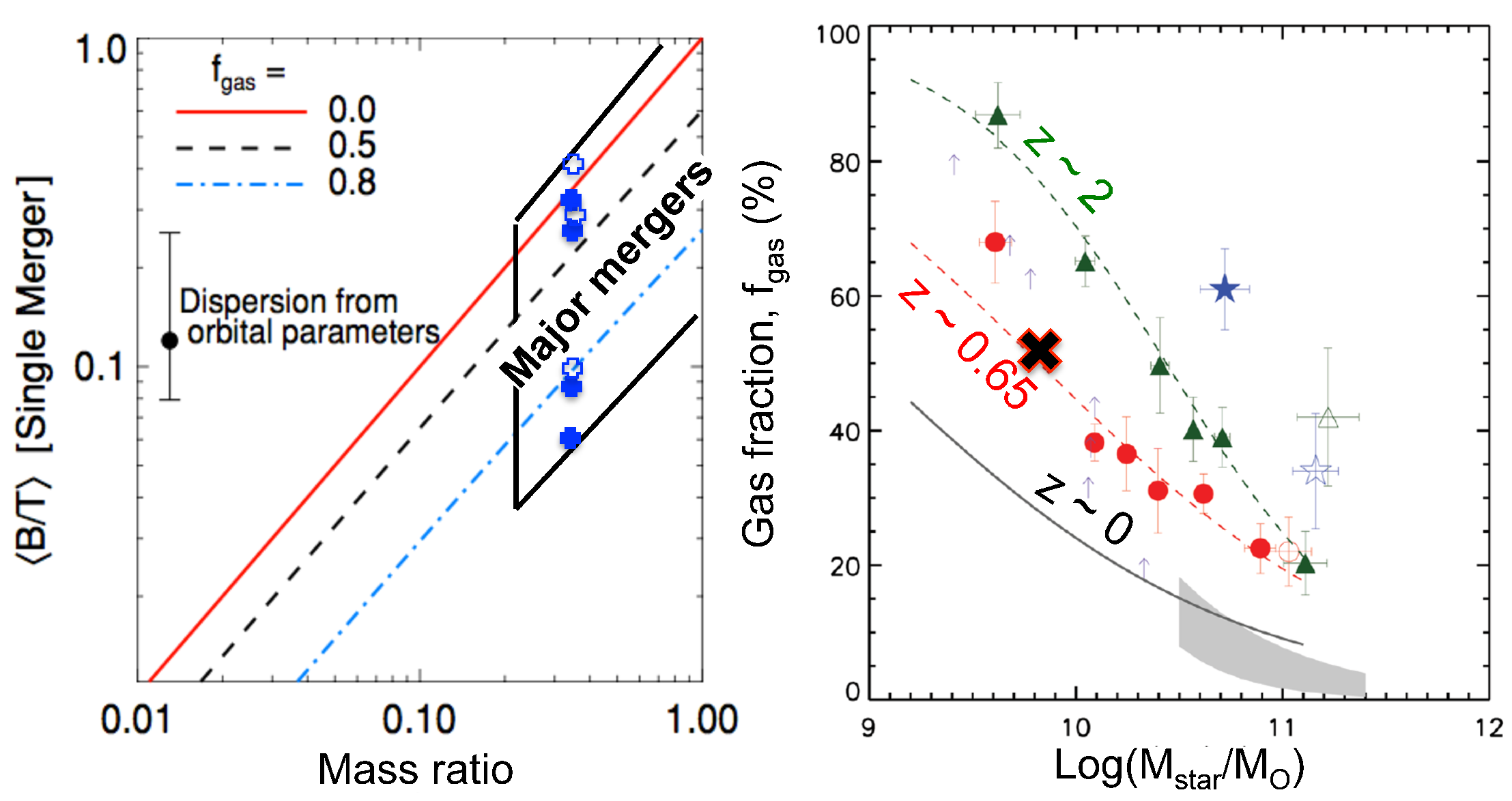}}
\caption{\footnotesize
A montage of Figure 7 from \cite{Hopkins10} (left) and Figure 7 from \cite{Rodrigues12} (right). On the left is shown the average B/T resulting from a merger, with 3  sampled gas fractions. The scatter between different orbits is very large, illustrated by an error bar, and also by 8 simulations (open and filled blue dots: constant and epoch-varying feedback, respectively) done by \cite{Hammer12} of MW-mass galaxies reformed after a 8 billion years old, 3:1 merger of 60\% gas-rich galaxies. On the left panel is shown the gas fraction - stellar mass relationship at 3 redshifts (points \& dashed lines), corresponding to look-back times of 10, 6 and 0 billion years, respectively. The cross indicates a median IMAGES galaxy progenitor at z= 0.83 (7 billion years ago) with 7.5 $10^{9} M_{\odot}$ and 50\% gas-richness (from \citet{Hammer09}), such galaxies being expected to be very common 7 to 10 billion years ago. This allows to form a large variety of spirals, possibly similar to the observed Hubble sequence.
}
\label{eta}
\end{figure*}

\section{The IMAGES survey and results}

A full and detailed description of the IMAGES survey can be found in \cite{Hammer12} and references therein. Here I just underline why this survey is the most relevant for studying the progenitors of present day spiral galaxies:
\begin{enumerate}
\item Completeness \& representativeness: galaxies are selected from large and complete sample of galaxies \citep{Ravikumar07} with a single absolute magnitude in J band ($M_{J}(AB)<$ -20.3), warranting a selection as closest as possible linked to stellar mass; 
\item Conservative \& robust morphological classification \citep{Delgado10}: 4-color images from HST/ACS are available and a procedure using (1) rest-frame wavelength, depth and PSF equivalent to that of SDSS local galaxies; (2)  calculation of half light radius to eliminate object too compact for being classified; (3) light decomposition from GALFIT software (Peng et al.) to deliver B/T, residuals and scale lengths of bulge and disk, (4) a color map allowing to identify low S/N in the outskirts from the method of \cite{Zheng04}, and (5) a decision tree allowing a robust classification remarkably similar to that of a recognized world expert \citep{vandenBergh02}.
\item Complimentary spatially resolved kinematics and well resolved morphologies from HST are available for all IMAGES galaxies; it is only with this knowledge that one can robustly classify the dynamical status of a galaxy leading \cite{Hammer09} to introduce the morpho-kinematics classification.
\end{enumerate}
These procedures ensures the representativeness of IMAGES galaxies, which are equivalent to the progenitors of $M_{stellar}>$ 1.5 $10^{10}$ $M_{\odot}$ galaxies, according to the Cosmological Principle. The conservative morphological classification is essential because it correlates almost perfectly with kinematical classification: peculiar kinematics are associated to peculiar morphologies at the level of 90\% \citep{Neichel08}. Such a correlation between morphological features linked to stars and kinematical features linked to the ionized gas clearly indicates a fundamental physical process affecting the whole galaxy formation.

Besides IMAGES (z$\sim$ 0.65), other spatially resolved kinematics surveys are providing similar qualitative results, though quantitatively they could be only indicative, because:
\begin{enumerate}
\item they are lacking of precise selection criteria \citep[SINS z$\sim$ 2,][]{Forster-Schreiber06,Forster-Schreiber09} or are selected with an apparent magnitude coinciding with a rest-frame at UV wavelengths \citep[MASSIV z$\sim$1.5, ][]{Contini12}, leading to an important contamination by bright starbursts in dwarves; 
\item they don't have or have only sparse imagery from HST preventing a robust kinematical classification. For example, to warrant the identification of a rotating disk requires to verify whether or not the dynamical axis coincides with the main axis of a spiral.
\end{enumerate}
 Thus IMAGES is still the unique sample allowing to study the detail physical properties of spiral galaxy progenitors.
It results that $\sim$ 50\% of spiral progenitors have peculiar morpho-kinematics properties 6 billion years ago, which prevent them to be classified in the Hubble Sequence \citep{Hammer09}.  The discrepancy between present-day spirals and their progenitors is not trivial: peculiar galaxies are entirely responsible of the large scatter of the Tully-Fisher relation at moderate redshift \citep{Flores06,Puech10}, which contrasts significantly with the tight relationship drawn by nearby spirals. Minor mergers are unlikely to induce such large angular momentum changes as well as affecting velocity fields on large scales. Outflows provoked by stellar feedback are not observed in the IMAGES galaxies that are generally forming stars at moderate rate \citep{Rodrigues12}. Internal fragmentation is limited because less than 20\% of the IMAGES galaxies show clumpy morphologies \citep{Puech10b} while associated cold gas accretion tends to vanish in massive halos at z$<$1, with rates $<$1.5 $M_{\odot}$/yr at
z$\sim$ 0.7 \citep{Keres09}. Finally perturbations from secular and internal processes (e.g. bars or spirals) are too small to be detected by the "large-scale" spatially resolved spectroscopy of IMAGES. Major mergers appear to be the most likely mechanism explaining abnormal morphologies as well as peculiar large scale motions of the gas in anomalous galaxies \citep{Hammer09,Hammer12,Puech12}. 

\section{Are all giant galactic disks formed through major mergers?}
Approximately half of the progenitors of spiral galaxies were in a major merger phase, 6 billion years ago, and their IMAGES counterparts have been successfully modeled \citep{Hammer09} using a mass ratio\footnote{These mass ratio are robust because they are observed in half of the cases, those before the fusion.} smaller than 4:1. Is it a too large fraction? In a detailed comparison between observations and theoretical expectations from a semi-analytic $\Lambda$CDM model  \citep{Hopkins10}, \cite{Puech12} have shown that both are in good agreement without any fine tuning. Such a high fraction is caused by the remarkable sensitiveness of the morpho-kinematics analysis to detect all the various merger phases, from the first distortions due to the first passage, to the perturbations during the rebuilding of the disk. Durations of such phases is generally 3-4 billion years for a MW-mass galaxy. An example of a rebuilding disk is well illustrated by J033245.11-274724.0 at z=0.4, which resembles to a typical dust-enshrouded  disk observed few hundred millions years after the fusion \citep{Hammer09b}. Such systems are very hard to recover in imagery, the dusty disk requiring 120 hours of HST/ACS (UDF) to be detected!

A significant fraction of giant spirals have had their disk rebuilt after a major merger episode, which may have occurred up to 10-12 billion year ago. This is now in agreement with the results of most (if not all) cosmological simulations \citep{Font11,Brook11,Guedes11,Keres12,Aumer13}. Do all spiral experienced such a process to form their disks? Perhaps a counter-argument is provided by bulge-less galaxies or by the numerous spiral galaxies with pseudo-bulges \citep[and references therein]{Kormendy13}, supposedly related to secular evolution conversely to classical bulges. This may cause a certain tension because large late type spirals are those requiring the largest angular momentum to support their disks. Simulations of realistic disks have been attempted, by using specific feedback mechanisms preventing star-formation in the core of the merger remnant \citep{Brook11}, though such prescriptions could appear somewhat ad-hoc. A better understanding of star-formation and feedback requires to account for all stellar processes, not only supernovae \citep{Hopkins13}.

Forming bulge less galaxies is expected through ancient gas-rich mergers, perhaps limited to favorable orbits.
 In Figure 1, open (constant feedback) and full (varying feedback with epoch) dots indicate the location of the 3:1, 60\% gas rich mergers from \cite{Hammer12} with 4 orbits (inclined, direct, retrograde and polar) defined by \cite{Barnes02}. These mergers have been let evolved during 8 billion years, leading to very realistic disks, most of them showing bulge with low Sersic index, and several being dominated by bars. It results that ancient gas-rich mergers often lead to the formation of pseudo-bulges \citep{Keselman12,Hammer12}. The  remarkable simulation of NGC5907 and of its loop system by \cite{Wang12}, is also that of a bulge less spiral galaxy, 8-9 billion years after the fusion. 

\section{Discussion \& conclusion}
In 2004, when I introduced the rebuilding disk scenario at the "30Doradus to Lyman-break galaxies" Conference in Cambridge, reactions were more than mitigated. This scenario was considered to be not consistent with our knowledge of the Milky Way, and so indeed the MW is apparently exceptional by having no significant merger during 10-11 billion years \citep{Hammer07}. Later on, it has been misinterpreted as being at odds with the  $\Lambda$CDM, while it indeed solves the disk formation or survival problem with a merger rate of $\sim$ one per L* galaxy during a Hubble time \citep{Stewart09,Puech12}. Recently, the star formation - stellar mass (SFR-$M_{star}$) relationship has been claimed to be too tight for allowing an important role for mergers \citep{Elbaz11}. Later relationship is however not free of numerous biases, e.g., selecting from far-IR is eliminating quiescent galaxies and artificially reducing the actual scatter. More importantly star formation during various merger phases always shows one or two peak(s), one of them at the fusion time \citep{Hammer05, Puech12,Hopkins13}, i.e., in a perfect agreement with galaxies having a "normal" activity for most of the time, interrupted by a short burst. Interestingly in \cite{Elbaz11} most star bursting galaxies that are off the SFR-$M_{star}$ relation are compact, such a result being already predicted by \cite{Hammer05} to probe the fusion, star bursting phase (see their Figure 6).

An important ingredient in modeling galaxy formation is the fraction of baryonic matter, a value that decreases from $\sim$ 20\% \citep[]{Barnes02} to less than 5\% in recent simulations \citep{Brook11,Hopkins13}, while it has been kept at 20\% for reproducing detailed observations of galaxies \citep{Hammer10,Wang12}. Low baryonic content is assumed through the halo matching technique, which is not proven to be unique, and possibly at odds with individual object studies. For example, The Large Magellanic Cloud dynamical mass within a 9 kpc radius is 4 times larger than its baryonic mass \citep{van der Marel09}. The halo matching technique would predict a ratio larger than 60 and let the overall LMC mass (1.8 $10^{11} M_{\odot}$) passing the total mass of the truncated MW disk (0.55 $10^{11} M_{\odot}$)! Observers of the Southern Hemisphere might need to be convinced.
Both low and high dark-to-baryonic matter ratio can be accommodated to predict disk rebuilding after mergers. However the fraction of baryonic matter ejected from a merger depends closely on this ratio, leading to low values in a very dark-matter dominated system.  Perhaps this explains why only the feedback mechanism is considered for estimating how many baryons have been expelled in the intergalactic medium \citep{Brook11,Hopkins13}. Mergers could be also the culprit of a large fraction of missing baryons \citep{Hammer12}, under the hypothesis that galactic baryonic fractions are not negligible when compared to $\Lambda$CDM values \citep[17.5\%, ][]{Komatsu09}. 

If most galaxies have experienced a major merger, it should have left fossil imprints into their halo. The numerous, low-surface brightness stellar streams discovered in
the outskirts of nearby, isolated spiral galaxies \citep{Martinez-Delgado10} are likely relics of these ancient major mergers. Recent minor mergers are unlikely to be at their origin because
of the stream red colors and the absence of any residual core. The whole NGC5907
galaxy (disk, bulge and thick disk) and associated loops have been successfully
modeled by assuming a 3:1 gas-rich major merger 8 to 9 billion years ago \citep{Wang12}.
Our nearest neighbour, M31
shows a classical bulge and a high halo metallicity suggesting a major merger
origin \citep{vandenBergh05,Kormendy13}. 
The later provides a robust explanation of the stellar Giant Stream, which could be made of
tidal tail stars captured by the galaxy gravitational potential after the fusion time \citep{Hammer10}.
In fact stars of the Giant Stream have ages older than 5.5 Gyr, which is difficult to
reconcile with a recent minor merger. A 3:1 gas-rich merger may reproduce the
M31 substructures (disk, bulge \& thick disk) as well as the Giant Stream assuming
the interaction and fusion may have occurred 8.75$\pm$0.35 and 5.5$\pm$0.5 Gyr ago,
respectively.

Perhaps surprisingly, the merger model of M31 of \cite{Hammer10} predicted the disk of satellites recently discovered by \cite{Ibata13}. Indeed, 3D locations of these M31 satellites are drawing two loop systems \citep{Hammer13} similar to those found around NGC5907, i.e., typical signatures of ancient major mergers. The fact that both the MW and M31 have a disk of satellites have lead \cite{Hammer13} to suggest that they are relics of the ancient M31 merger. The two galaxies and their fossil systems might be linked together through the modeled orbital angular plane and the \cite{Ibata13} disk of satellites because both are pointing to the MW. If true, many Local Group dwarves could be relics of tidal dwarves, without dark matter. 
A better understanding of galaxy formation will certainly bring us with many surprises in the future.


\bibliographystyle{aa}

\begin{thebibliography}{}
\bibitem[Aumer et al.(2013)]{Aumer13}Aumer, M., White, S. D. M., Naab, T., Scannapieco, C., 2013 MNRAS, 434, 3142
\bibitem[Barnes(2002)]{Barnes02} Barnes, J.E., 2002 MNRAS, 333, 481 
\bibitem[Brook et al.(2011)]{Brook11} Brook, C. B.; Stinson, G.; Gibson, B. K. et al., 2011  MNRAS,  419, 771
\bibitem[Contini et al.(2012)]{Contini12} Contini, T. et al. 2012, A\&A, 539, 91
\bibitem[Danovich et al.(2012)]{Danovich12} Danovich, M., Dekel, A., Hahn, O. and Teyssier, R., 2012, MNRAS, 422, 1732 
\bibitem[Delgado-Serrano et al.(2010)]{Delgado10} Delgado-Serrano, R.; Hammer, F.; Yang, Y. B.et al., 2010 A\&A, 509, 78 
\bibitem[Elbaz et al.(2011)]{Elbaz11} Elbaz, D. et al., 2011 \aap 533, 119
\bibitem[Flores et al.(2006)]{Flores06} Flores, H., Hammer, F., Puech,  M.et al., 2006 A\&A, 455, 107 
\bibitem[Font et al.(2011)]{Font11} Font, A. S., McCarthy, I. G., Crain, R. A. et al., 2011 MNRAS 416, 2802
\bibitem[Forster-Schreiber et al.(2006)]{Forster-Schreiber06} Forster-Schreiber, N. M., Genzel, R., Lehnert. M. et al., 2006, \apj,  645, 1062
\bibitem[Forster-Schreiber et al.(2009)]{Forster-Schreiber09} Forster-Schreiber, N. M., Genzel, R., Bouch«e, N. et al., 2009, \apj,  706, 1364
\bibitem[Guedes et al.(2011)]{Guedes11} Guedes, J., Callegari, S., Madau, P. et al., 2011 \apj,  742, 76
\bibitem[Hammer et al.(2005)]{Hammer05} Hammer, F., Flores, H.,  Elbaz, D., Zheng, X.~Z., Liang, Y.~C., \& Cesarsky, C.  2005, A\&A,  430, 115 
\bibitem[Hammer et al.(2007)]{Hammer07} Hammer, F., Puech, M.,  Chemin, L. et al., 2007 \apj, 662, 322
\bibitem[Hammer et al.(2009)]{Hammer09} Hammer, F., Flores, H., Puech, M., Yang, Y.~B., Athanassoula, E., et al.\ 2009, A\&A, 507, 1313
\bibitem[Hammer et al.(2009b)]{Hammer09b} Hammer, F., Flores, H., Yang, Y.~B., Athanassoula, E., et al.\ 2009b, A\&A, 496, 381
\bibitem[Hammer et al.(2010)]{Hammer10} Hammer, F., Yang, Y.~B., Wang, J.~L. et al. \ 2010, ApJ, 725, 542 
\bibitem[Hammer et al.(2013)]{Hammer13} Hammer, F., Yang, Y.~B., Fouquet, S. et al. \ 2013, MNRAS, 431, 2543 
\bibitem[Hammer et al.(2012)]{Hammer12} Hammer, F., Yang, Y.~B., Flores, H., Puech, M. 2012, Modern Physics Letters A, 33, 1230034 (arXiv:1110.1376) 
 \bibitem[Ibata et al.(2013)]{Ibata13} Ibata, R., Lewis, G., Conn, A. R. et al. \ 2013, Nature, 493, 62 
\bibitem[Hopkins et al.(2010)]{Hopkins10} Hopkins, P. F., Bundy, K., Croton, D. et al., 2010, ApJ., 715, 202
\bibitem[Hopkins et al.(2013)]{Hopkins13} Hopkins, P. F., Keres D., Onorbe, J. et al., 2013, (arXiv:1311.2073)
\bibitem[Keres et al.(2009)]{Keres09} Keres, D., Katz, N., Fardal, M. et al., 2009 MNRAS, 395, 160
\bibitem[Keres et al.(2012)]{Keres12} Keres, D., Vogelsberger, M., Sijacki, D. et al., 2012 MNRAS, 425, 2027 
\bibitem[Keselman \& Nusser(2012)]{Keselman12} Keselman, J. A., Nusser, A., 2012 MNRAS, 424, 1232
\bibitem[Komatsu et al.(2009)]{Komatsu09} Komatsu, E., et al. 2009 ApJ S.S., 180, 330
\bibitem[Kormendy(2013)]{Kormendy13} Kormendy, J. 2013, In Secular Evolution of Galaxies, XXIII Canary Islands Winter School of Astrophysics, ed. J. Falc\'on-Barroso \& J. H. Knapen (Cambridge: Cambridge Univ. Press) (arXiv:1311.2609)
\bibitem[Maller, Dekel, \& Somerville(2002)]{Maller02} Maller A.~H., Dekel A., Somerville R., 2002, MNRAS, 329, 423
\bibitem[Martinez-Delgado et al.(2010)]{Martinez-Delgado10} Mart{\'{\i}}nez-Delgado, D., et al., 2010, A. J., 140, 962 
 \bibitem[Neichel et al.(2008)]{Neichel08} Neichel, B., et al.\ 2008, A\&A, 484, 159 
\bibitem[Puech(2010)]{Puech10b} Puech M., 2010 MNRAS,  406, 535 
\bibitem[Puech et al.(2010)]{Puech10}Puech, M., Hammer, F.; Flores, H., Delgado-Serrano, R., Rodrigues, M.,Yang, Y., 2010, A\&A 510, 68
 \bibitem[Puech et al.(2012)]{Puech12} Puech, M., Hammer, F., Hopkins, P. F., Athanassoula, E., Flores, H., Rodrigues, M., Wang, J. L. \ 2012, ApJ, 753, 128
\bibitem[Ravikumar et al.,(2007)] {Ravikumar07} Ravikumar, C. D. et al., 2007, \aap, 465, 1099
\bibitem[Rodrigues et al.(2012)]{Rodrigues12} Rodrigues, M., Puech, M., Hammer, F. et al., 2012 MNRAS, 421, 2888 
\bibitem[Steinmetz \& Navarro(1999)]{Steinmetz99}Steinmetz, M. \& Navarro, J. 1999, \apj, 513, 555
\bibitem[Stewart et al.(2009)]{Stewart09} Stewart, K., \& Bullock, J.~S., Wechsler, R. et al., 2009, \apj, 702, 307 
\bibitem[Toomre \& Toomre(1972)]{Toomre72} Toomre, A., Toomre, J., 1972, {\it Ap. J.}, 178, 623
 \bibitem[van den Bergh(2002)]{vandenBergh02} van den Bergh S., 2002, Pub. Astron. Soc. Pacific , 114, 797 
\bibitem[van den Bergh(2005)]{vandenBergh05} van den Bergh, S., 2005, in {\it The Local Group as an Astrophysical Laboratory}, ed. M. Livio\& T. M. Brown (Cambridge: Cambridge Univ. Press), P.1-15  
\bibitem[van der Marel et al.(2009)]{van der Marel09} van der Marel, R. P., Kallivayalil, N., Besla, G., 2009, Proc. IAU Symposium, Vol. 256, p. 81-92
 \bibitem[Wang et al.(2012)]{Wang12} Wang, J., Hammer, F., Athanassoula, E., Puech, M., Yang, Y., et al.\ 2012, A\&A, 538, A121
\bibitem[Yang et al.(2008)]{Yang08} Yang, Y., Flores, H., Hammer, F., et al. 2008, A\&A, 477, 789
\bibitem[Zheng et al.(2004)]{Zheng04} Zheng X.~Z., Hammer
  F., Flores H., Ass{\'e}mat F., Pelat D., 2004, A\&A, 421, 847

\end{thebibliography}

\end{document}